\documentclass[prl,twocolumn,showpacs,amsmath,amssymb,superscriptaddress]{revtex4-2}
\usepackage{epsfig,epstopdf} 
\usepackage{graphicx}
\usepackage{dcolumn}
\usepackage{bm}
\usepackage{bbm}
\usepackage{ulem}
\usepackage{xcolor}
\usepackage{amsmath}
\usepackage{esint}
\usepackage{slashed}
\usepackage{mathrsfs}
\usepackage{subfigure}
\usepackage{verbatim}
\usepackage{dsfont}
\usepackage{float}
\usepackage{wasysym}

\usepackage[breaklinks]{hyperref}
\hypersetup{colorlinks=true, linkcolor=blue, citecolor=blue, filecolor=blue, urlcolor=blue}

\AtBeginDocument{%
	\newwrite\bibnotes
	\def\bibnotesext{Notes.bib}
	\immediate\openout\bibnotes=\jobname\bibnotesext
	\immediate\write\bibnotes{@CONTROL{REVTEX41Control}}
	\immediate\write\bibnotes{@CONTROL{%
			apsrev41Control,author="08",editor="1",pages="1",title="0",year="1"}}
	\if@filesw
	\immediate\write\@auxout{\string\citation{apsrev41Control}}%
	\fi
}%

\begin{document}

\bibliographystyle{apsrev4-1}

\title{Intrinsic Time-reversal-invariant Topological Superconductivity in Thin Films of Iron-based Superconductors}
\author{Rui-Xing Zhang}
\email{ruixing@umd.edu}
\author{S. Das Sarma}
\affiliation{Condensed Matter Theory Center and Joint Quantum Institute, Department of Physics, University of Maryland, College Park, Maryland 20742-4111, USA}

\begin{abstract}
	We establish quasi-two-dimensional thin films of iron-based superconductors (FeSCs) as a new high-temperature platform for hosting intrinsic time-reversal-invariant helical topological superconductivity (TSC). Based on the combination of Dirac surface state and bulk extended $s$-wave pairing, our theory should be directly applicable to a large class of experimentally established FeSCs, opening a new TSC paradigm. In particular, an applied electric field serves as a ``topological switch" for helical Majorana edge modes in FeSC thin films, allowing for an experimentally feasible design of gate-controlled helical Majorana circuits. Applying an in-plane magnetic field drives the helical TSC phase into a higher-order TSC carrying corner-localized Majorana zero modes. Our proposal should enable the experimental realization of helical Majorana fermions.  
\end{abstract}

\maketitle

{\it Introduction} - The past years have witnessed remarkable experimental progress in revealing topological band structures in a variety of iron-based superconductors (FeSCs)~\cite{zhang2018observation,wang2018evidence,shi2017monolayer,liu2018robust,Zhang2019multiple,chen2019quantized,Machida2019zero,peng2019observation,Kong2019half,zhu2020nearly,Liu2020newMajorana,kong2020tunable}. As a prototypical example of topological FeSCs, FeTe$_{1-x}$Se$_x$ features a single helical Dirac surface state in its normal state~\cite{wang2015topological,xu2016topological,wu2016topological}. Below the superconducting transition temperature, such a Dirac surface state is fully gapped out by the ``self-proximity" effect of the bulk superconductivity~\cite{zhang2018observation}, becoming a surface topological superconductor (TSC) proposed by Fu and Kane~\cite{fu2008superconducting}. When a magnetic field is applied, strong evidence for vortex Majorana bound states has been reported in FeTe$_{1-x}$Se$_x$ by several experimental groups~\cite{wang2018evidence,Machida2019zero}. Notably, the vortex Majorana signature has also been observed in (Li$_{0.84}$Fe$_{0.16}$)OHFeSe~\cite{liu2018robust,chen2019quantized}, CaKFe$_4$As$_4$~\cite{Liu2020newMajorana}, LiFeAs~\cite{kong2020tunable}, etc. Meanwhile, the combination of band topology and unconventional pairing symmetry of FeSCs has motivated researchers to explore new topological aspects of FeSCs that go beyond the Fu-Kane paradigm, including the recent theoretical and experimental progress on higher-order topological phenomena~\cite{zhang2019helical,zhang2019higher,wang2018high,wu2019high,gray2019evidence,wu2020boundary}. 

On the other hand, realization of time-reversal-invariant (TRI) TSCs~\cite{qi2009time,fu2010odd,nakosai2012topological,zhang2013time,wang2014two,haim2019time,casas2019proximity} remains a decade-old outstanding question in the field of topological matter. In particular, little experimental progress has been made towards realizing two-dimensional (2D) TRI helical TSC~\cite{qi2009time,zhang2013time}, which features helical Majorana edge modes and manifests as the superconducting counterpart of quantum spin Hall effect (which itself can be thought of as the TRI version of the integer quantum Hall effect). As pointed out in Ref.~\cite{qi2010topological}, helical TSC occurs only when there exists an odd number of TRI-momentum-enclosing Fermi surfaces with a negative superconducting pairing function. To satisfy this stringent topological constraint, most theoretical proposals of helical TSC rely on proximity effects involving either unconventional superconductors~\cite{zhang2013time,haim2016nogo} or two $s$-wave superconductors with a precise $\pi$-phase difference~\cite{liu2011helical}, both of which are experimentally highly challenging. 

\begin{figure}[t]
	\centering
	\includegraphics[width=\columnwidth]{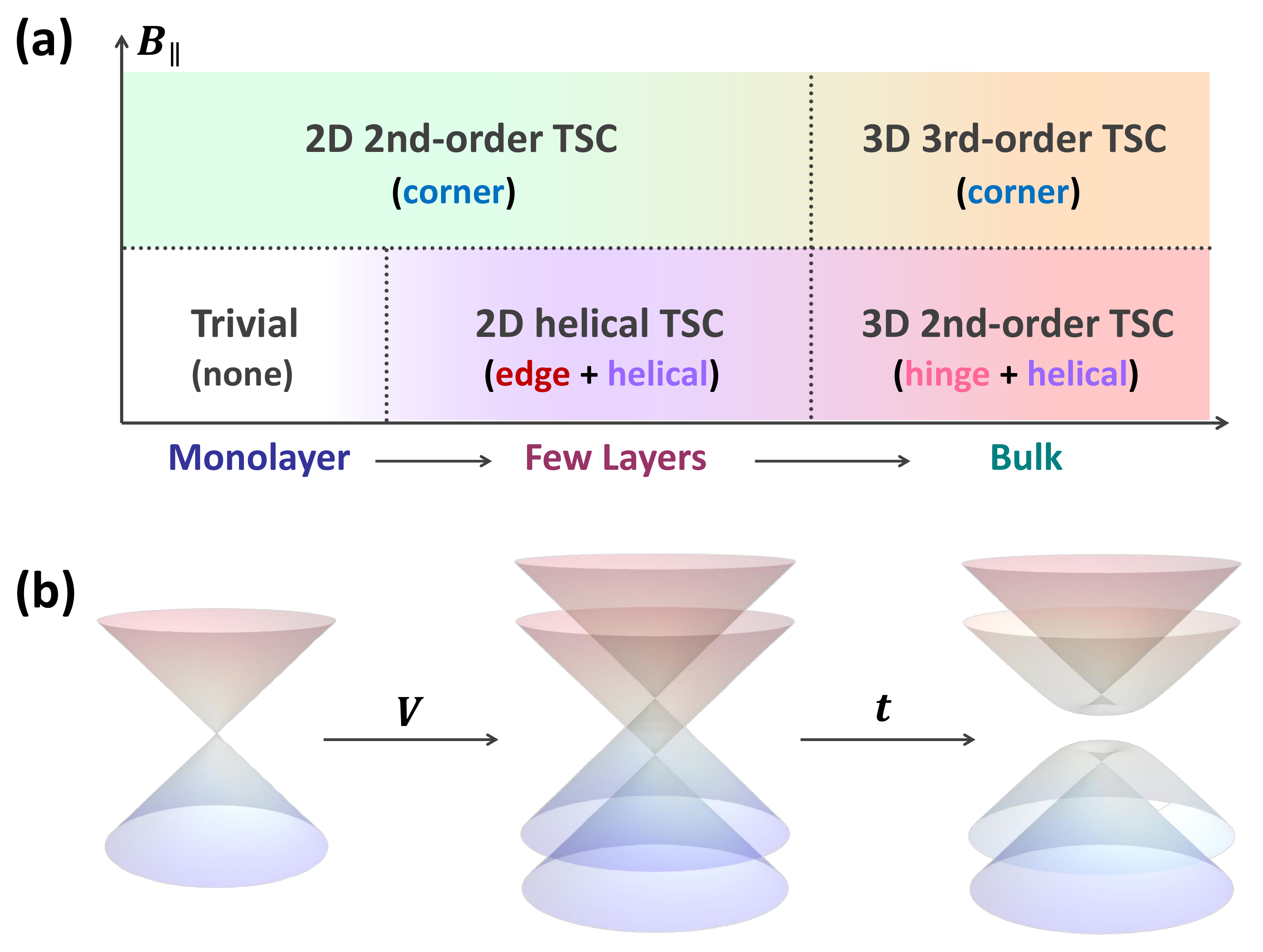}
	\caption{(a) Phase diagram of topological FeSCs as a function of sample thickness and an applied in-plane magnetic field ${\bf B}_\parallel$, which is based on \cite{zhang2019helical,zhang2019higher,wu2019high}. The characteristics of boundary Majorana modes for each phase are summarized. For example, ``hinge $+$ helical" implies 1D helical Majorana modes living on the hinges of a 3D 2nd-order TSC phase, while ``corner" indicates the Majorana corner modes for a 2D 2nd-order (or a 3D 3rd-order) TSC phase. (b) In a quasi-2D FeSC thin film, low-energy Rashba bands arise from (i) the hybridization $t$ between top and bottom Dirac surface states; (ii) the structural inversion asymmetry $V$ from an applied out-of-plane electric field. The formation of 2D Rashba bands is crucial for achieving helical TSC in FeSC thin films.}
	\label{Fig1}
\end{figure}

In this work, we propose quasi-2D thin films of FeSCs as a new paradigm for hosting {\it intrinsic} helical topological superconductivity (Fig.~\ref{Fig1}), which, unlike the previous proposals, does not require any external proximity effect. Based on two experimentally supported phenomena in FeSCs, i.e., the Dirac surface state and the bulk extended $s$-wave (i.e., $s_\pm$) pairing, we show that the structural inversion asymmetry (SIA)~\cite{shan2010effective,zhang2013electric} from an applied electric field is the key control knob for helical TSC phase in FeSC thin films, allowing for a remarkable electrical tunability of the emergent helical Majorana modes. The topological condition for helical TSC is identified through an effective theory analysis, which we verify by direct numerical calculations. By constructing a minimal lattice model for a 4-layer FeSC thin film, we confirm the topological nature of our desired helical TSC phase by calculating both its edge helical Majorana modes and the bulk $\mathbb{Z}_2$ topological index. We also systematically map out the topological phase diagram as a function of pairing parameters, chemical potential, and SIA term, which serves as a useful guide for searching and controlling Majorana modes in future experiments. While a helical TSC does not feature Majorana bound states on its own, applying an in-plane magnetic field necessarily drives the helical TSC into a 2D higher-order TSC with corner-localized Majorana zero modes.

{\it Effective Theory} - We start with an effective theory description of FeSC thin films to illustrate the topological condition for realizing the helical TSC phase. The normal state of FeSC thin films can be described by the Hamiltonian       
\begin{equation}
h_\text{eff} ({\bf k}) = v_F \rho_z \otimes (k_x s_y - k_y s_x) + t \rho_x  + V \rho_z - \mu.
\label{eq:effective Ham}
\end{equation}
The first term in Eq.~\ref{eq:effective Ham} comes from the Dirac surface states from both the top and bottom layers of the thin film, where the Pauli matrices $\rho$ and $s$ denote layer and spin degrees of freedom, respectively. In addition to the intersurface tunneling $t$ and the chemical potential $\mu$, an applied electric field along the out-of-plane direction leads to a potential difference between the top and bottom surfaces, which is captured by the SIA term $V$ in Eq.~\ref{eq:effective Ham}. As schematically demonstrated in Fig.~\ref{Fig1} (b), the SIA explicitly breaks the spatial inversion symmetry and drives the Dirac surface states into a set of 2D Rashba bands~\cite{zhang2013electric}. In particular, the eigenspectrum of $h_\text{eff}({\bf k})$ is analytically solvable with $E = \pm \sqrt{(vk\pm V)^2 + t^2}-\mu$, where $k=\sqrt{k_x^2+k_y^2}$. 
When $\mu>t>0$, the Rashba nature of $h_\text{eff}({\bf k})$ directly indicates the existence of two Fermi surfaces, denoted as $FS^{(+)}$ and $FS^{(-)}$. For $V>0$, the Fermi momentum of $FS^{(\pm)}$ is given by
\begin{eqnarray}
k_F^{(\pm)} &=& \frac{\sqrt{\mu^2-t^2}\pm V}{v_F}\ \ \text{for } \mu>\sqrt{t^2+V^2}, \nonumber \\
k_F^{(\pm)} &=& \frac{V\pm \sqrt{\mu^2-t^2}}{v_F}\ \ \text{for } t<\mu<\sqrt{t^2+V^2}.
\end{eqnarray}

The bulk $s_\pm$-wave pairing $\Delta({\bf k})=\Delta_0+\Delta_1(\cos k_x + \cos k_y)$ provides a self-proximity effect on the Dirac surface states. Since the pairing function is expected to flip its sign while evolving from the Brillouin zone (BZ) center to the BZ corner, we require $|\Delta_0|<2\Delta_1$ for $\Delta_1>0$. The momentum-dependent pattern of pairing function requires $\Delta({\bf k})$ to vanish along a closed loop in the 2D BZ. In particular, by expanding $\Delta({\bf k})$ around the BZ center, such a zero-pairing loop with $\Delta({\bf k})=0$ can be approximated by a ring with a radius of
\begin{equation}
	k_\Delta = \sqrt{2(\frac{\Delta_0}{\Delta_1}+2)}.
\end{equation} 
Notably, the helical TSC phase emerges when $FS^{(+)}$ and $FS^{(-)}$ experience pairing potentials with opposite signs~\cite{zhang2013time}. Therefore, the topological condition for helical TSC is fulfilled when
\begin{equation}
k_F^{(-)}<k_\Delta < k_F^{(+)}.
\label{eq:topological condition}
\end{equation}
Obviously, there are enough tuning parameters in the problem to satisfy this condition.

\begin{figure}[t]
	\centering
	\includegraphics[width=0.48\textwidth]{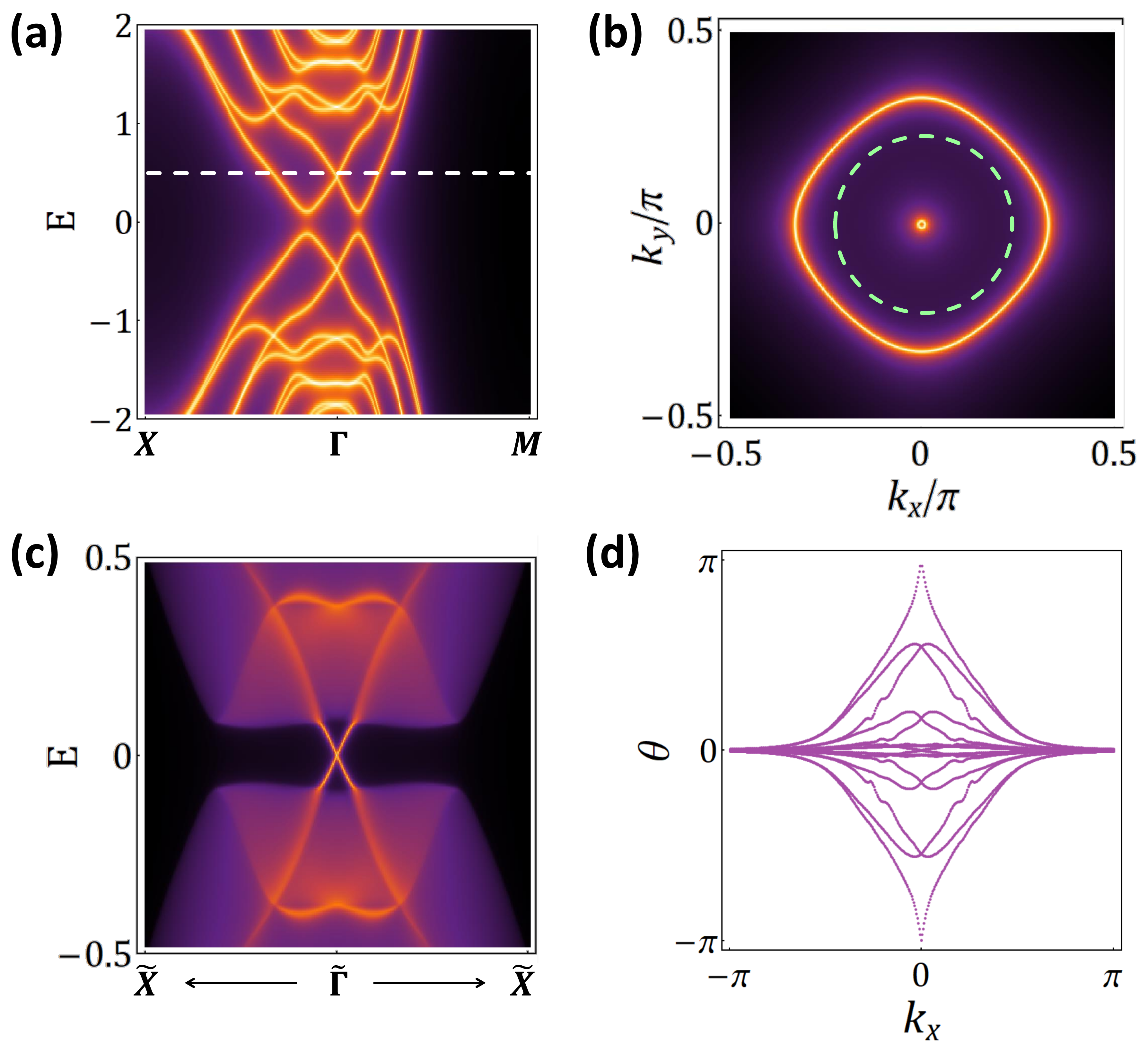}
	\caption{(a) Normal-state energy spectrum for a quasi-2D four-layer thin film, where the 2D Rashba bands are essentially hybridized Dirac surface states. The white dashed line denotes the Fermi level $\mu=0.5$. (b) Fermi surface plot at $\mu=0.5$. The green dashed circle denotes where the pairing function $\Delta({\bf k})$ vanishes. (c) Edge spectrum shows a pair of helical edge Majorana modes in the helical TSC phase. (d) Wilson loop spectrum $\theta$ features a gapless helical winding pattern, indicating the $\mathbb{Z}_2$ index $\nu=1$.}
	\label{Fig2}
\end{figure}

{\it Lattice Model and Helical TSC} - We now proceed to numerically confirm the helical TSC phase in a minimal lattice model. The normal-state topological physics of a 3D bulk FeSC is captured by a 4-band model on a cubic lattice, whose Hamiltonian in momentum space is given by $H_0({\bf k}) = v_k (\sin k_x \gamma_1 + \sin k_y \gamma_2 + \sin k_z \gamma_3)+M(k) \gamma_5$ with $M(k) = M_0-M_1(\cos k_x + \cos k_y) - M_2 \cos k_z$ \cite{zhang2019helical}. We have defined $\gamma_1= \sigma_x\otimes s_x, \gamma_2 = \sigma_x\otimes s_y, \gamma_3 = \sigma_x\otimes s_z, \gamma_4 = \sigma_y \otimes s_0, \gamma_5 = \sigma_z\otimes s_0$, where the Pauli matrices $\sigma_{0,x,y,z}$ denote the orbital degree of freedom. In particular, the spatial inversion symmetry is ${\cal I}=\gamma_5$ and the time-reversal operation is $\Theta=is_y{\cal K}$, where ${\cal K}$ denotes the complex conjugation. When $v=1,M_0=-4,M_1=-2,M_2=2$, $H_0({\bf k})$ has a single band inversion at $Z=(0,0,\pi)$ between energy bands with opposite inversion parities, capturing the key feature of the topological bands discovered in FeTe$_{1-x}$Se$_{x}$ and other topological FeSCs. According to the Fu-Kane parity criterion~\cite{fu2007topological}, $H_0({\bf k})$ with our choice of parameters describes a strong TRI TI with a single gapless 2D Dirac mode on its every surface. 

The normal state of a topological FeSC thin film can thus be modeled by constructing a $N_z$-layer slab with $H_0({\bf k})$, where $N_z=4$ is taken in our thin film simulations. The SIA term is taken into account by including an onsite potential $\pm V/2$ to the top and bottom layer of the slab, respectively. For $V=1$, the energy spectrum of the FeSC thin film is shown in Fig.~\ref{Fig2} (a), where the SIA-induced large Rashba band splitting is clearly revealed around $\Gamma$ point. In Fig.~\ref{Fig2} (b), we map out the Fermi surface for $\mu=0.5$ [the white dashed line in Fig.~\ref{Fig2}~(a)] and confirm the coexistence of $FS^{(+)}$ and $FS^{(-)}$ that are expected from the effective theory. This SIA-induced 2D Rashba band serves as the starting point for achieving the helical TSC phase.

We now explore the topological consequence of the intrinsic unconventional superconductivity in FeSCs by considering a Bogoliubov-de Gennes (BdG) Hamiltonian
\begin{equation}
	H_\text{BdG}({\bf k}) = \begin{pmatrix}
	H_\text{TF}({\bf k}) - \mu & -i \Delta({\bf k}) s_y \\
	i \Delta({\bf k}) s_y & -H_\text{TF}(-{\bf k})^T + \mu \\
	\end{pmatrix},
	\label{eq:BdG Hamiltonian}
\end{equation}
where $H_\text{TF}({\bf k})$ denotes the normal-state Hamiltonian for the 4-layer thin film. For the $s_\pm$-wave pairing function $\Delta({\bf k})=\Delta_0+\Delta_1(\cos k_x + \cos k_y)$, we take $\Delta_0=-0.7, \Delta_1=0.4$ so that the zero-pairing loop [shown by the green dashed line in Fig.~\ref{Fig2} (b)] exactly lies between $FS^{(\pm)}$, fulfilling the topological condition in Eq.~\ref{eq:topological condition}. We note that our parameter choices for achieving TSC are generic and not fine-tuned by any means.

To verify the helical topological nature of our system, we calculate the edge spectrum in a semi-infinite geometry with $k_x$ being a good quantum number. As shown in Fig.~\ref{Fig2} (c), a pair of 1D counterpropagating Majorana modes penetrates the bulk superconducting gap and crosses at exactly $\widetilde{\Gamma}$, a TRI momentum on the edge BZ. Therefore, the left and right-moving Majorana channels manifest themselves as a pair of time-reversal-related helical edge Majorana modes, the defining boundary characteristic for a helical TSC. We further extract its bulk $\mathbb{Z}_2$ topological index by calculating its Wilson loop spectrum. As shown in Fig.~\ref{Fig2} (d), the eigenspectrum $\theta$ of the Wilson loop operator exhibits a nontrivial helical winding pattern, which directly indicates a topologically nontrivial $\mathbb{Z}_2$ index $\nu=1$. Therefore, the edge helical Majorana modes and the bulk $\mathbb{Z}_2$ index $\nu=1$ together establish the helical TSC nature of our thin film, agreeing with the analytical Fermi-surface-based diagnosis from Eq.~\ref{eq:topological condition}.  

\begin{figure}[t]
	\centering
	\includegraphics[width=\columnwidth]{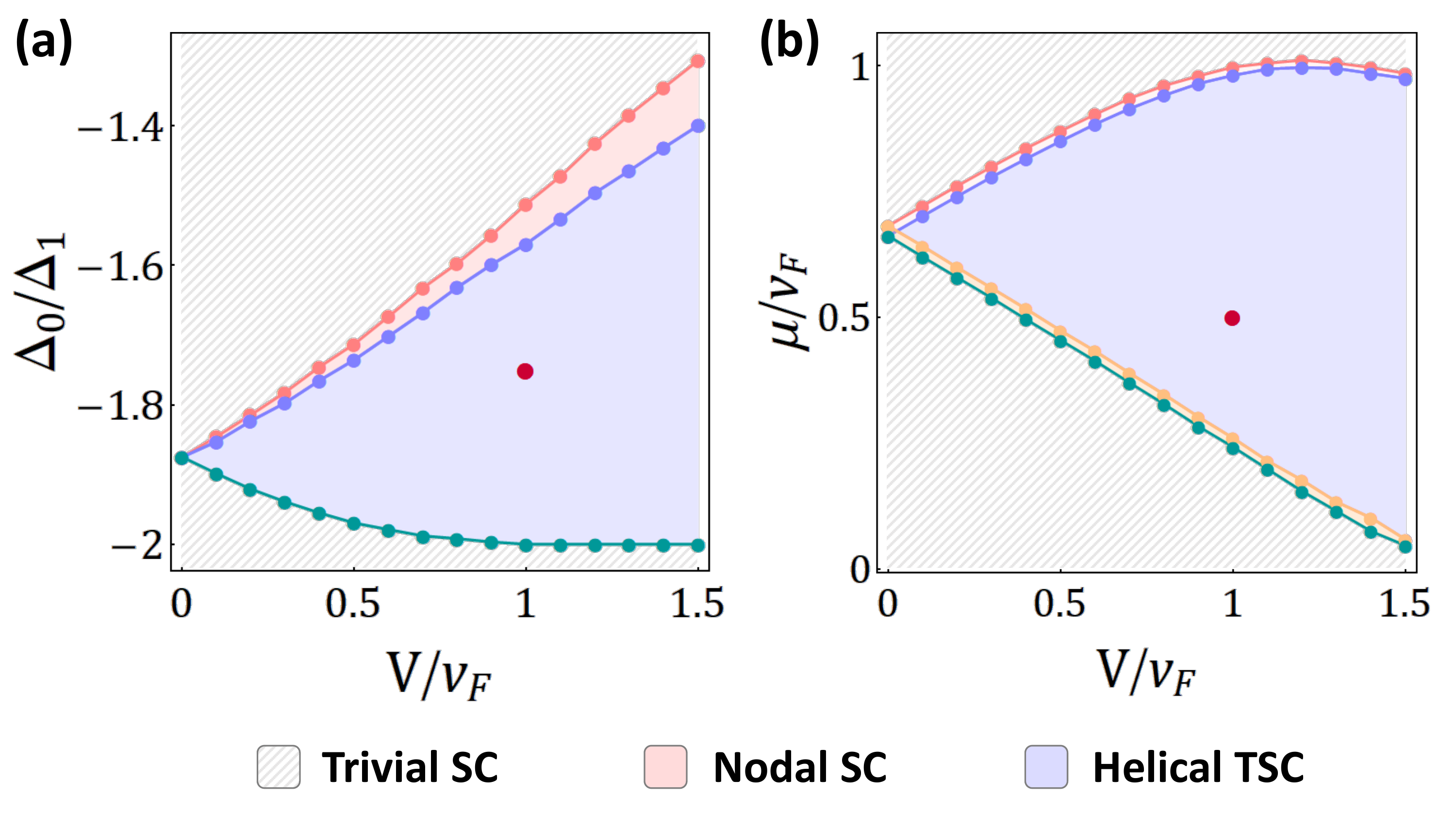}
	\caption{(a) Topological phase diagram for a fixed $\mu=0.5$. (b) Topological phase diagram for a fixed value of $\Delta_0/\Delta_1=-1.75$. The red dots in both (a) and (b) denote the location of the helical TSC phase shown in Fig.~\ref{Fig2}.}
	\label{Fig3}
\end{figure}

{\it Topological Phase Diagram} - The electrical tunability of both $\mu$ and $V$ in experiments motivates us to investigate the topological phase diagram for our FeSC thin film. In Fig.~\ref{Fig3} (a), we numerically map out the topological phase boundaries as a function of both the SIA term $V$ (in unit of $v_F$) and $\Delta_0$ (in unit of $\Delta_1$), for a fixed $\mu=0.5$. In particular, the blue fan-shaped region bounded by the blue and green dashed lines denotes the helical TSC phase, where the red dot labels the location of the parameter set of Fig.~\ref{Fig2} in the phase diagram. A similar fan-shaped region of helical TSC phase is also found in Fig.~\ref{Fig3} (b), where we plot the phase diagram by varying $\mu$ and $V$ (both in unit of $v_F$) while fixing $\Delta_0=-7/4\Delta_1$. The similar fan shape of the helical TSC regions in both Fig.~\ref{Fig3} (a) and (b) clearly illustrates the role of SIA in tuning the helical Majorana modes in FeSC thin films. In the Supplemental Material~\cite{supplementary}, we also map out the topological phase diagrams for thin films with $N_z=2$ and $N_z=6$, both of which resemble that of $N_z=4$ in Fig.~\ref{Fig3}. Therefore, we conclude that varying $N_z$ will only quantitatively modify our topological phase diagram, without introducing any qualitative change.

Besides the trivial superconducting phase labeled by the gray shaded region, we also uncover finite regions in both Fig.~\ref{Fig3} (a) and (b), where the BdG spectrum becomes nodal. Such a nodal superconducting state originates from the intersection between $FS^{(\pm)}$ and the zero-pairing loop, which serves as an intermediate phase between the helical TSC and trivial superconductors. We note that such a nodal state is absent in the effective theory based on Eq.~\ref{eq:effective Ham}, where both $FS^{(\pm)}$ and zero-pairing loop are approximated to be perfectly circular and hence cannot have point-like intersections. However, such an idealized circular assumption no longer holds in a real lattice system [see the Fermi surface plot in Fig.~\ref{Fig2} (b) as an example]. In particular, such doubly degenerate nodal points in class DIII admit a $\mathbb{Z}$ topological classification~\cite{schnyder2011topo} and can lead to Majorana flat bands lying at zero energy on certain edges. A detailed characterization of this nodal TSC phase is beyond the scope of the current work, and will be reported elsewhere. 

{\it Field-induced Higher-order Topology} - In the presence of an applied in-plane magnetic field ${\bf B}_{\parallel}$, the helical Majorana edge modes lose the protection from time-reversal symmetry, developing a Zeeman gap in the edge spectrum. Nonetheless, the edge Zeeman gap is necessarily anisotropic around the system boundary \cite{zhang2013surface}. For example, due to the spin-momentum locking, the right-moving Majorana channels for the top and bottom edges manifest exactly opposite spin textures and thus develop opposite edge Zeeman gaps. Consequently, when ${\bf B}_\parallel$ is not parallel to either $\hat{x}$ or $\hat{y}$ directions, the edge Zeeman gap will always be positive for half of the boundary, while being negative for the other half. Namely, applying a ${\bf B}_\parallel$ naturally generates a pair of mass domain walls for the edge helical Majorana modes when the Zeeman gap flips its sign, around which spatially localized Majorana zero modes should emerge.       

\begin{figure}[t]
	\centering
	\includegraphics[width=\columnwidth]{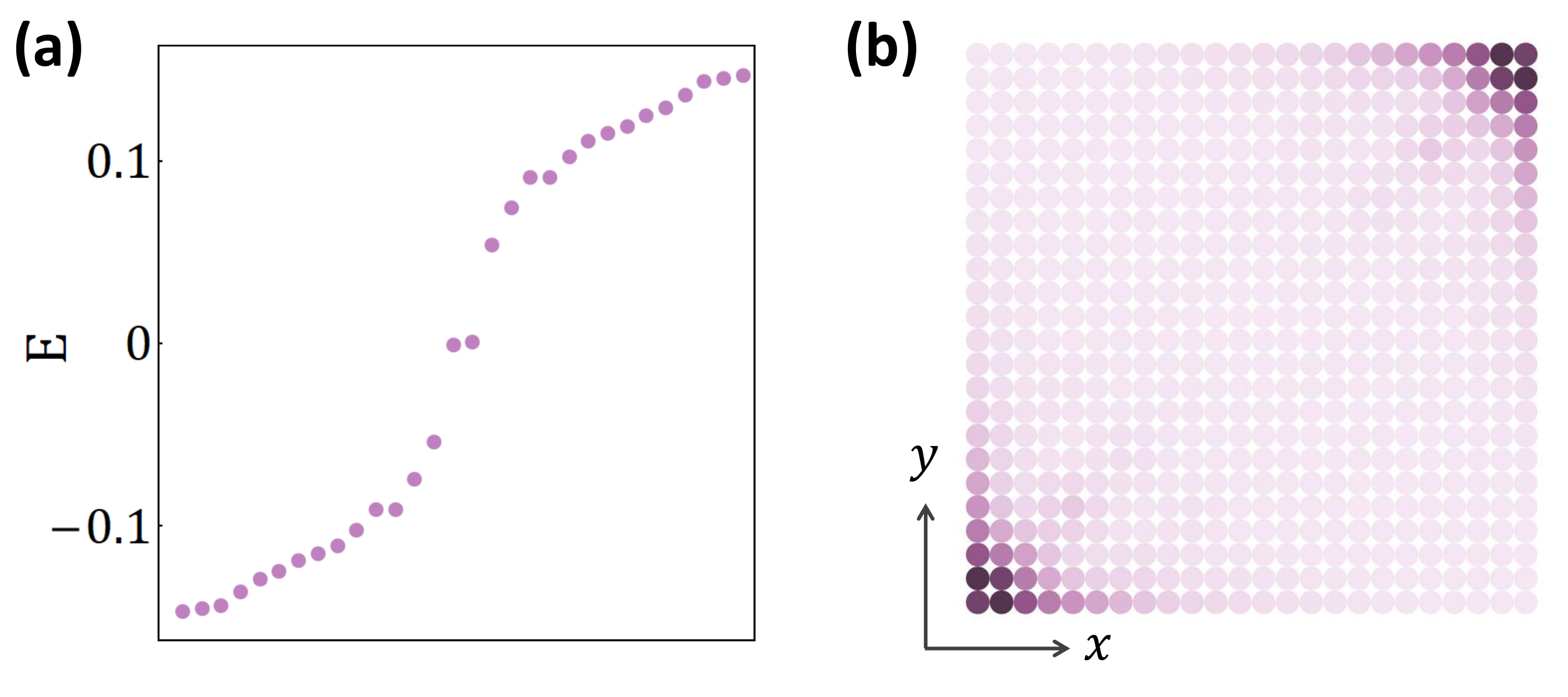}
	\caption{(a) Energy spectrum of the helical TSC phase when applying an in-plane magnetic field on a $24\times 24$ lattice. Despite the edge Zeeman gap, a pair of Majorana modes emerges at zero energy. (b) The spatial profile of the Majorana zero modes in (a) indicates their corner-localized nature.}
	\label{Fig4}
\end{figure}

To numerically confirm the above scenario, we include a bulk Zeeman term for our lattice model with
\begin{equation}
	H_\text{Zeeman}= g (\cos \varphi \tau_z\otimes \sigma_z \otimes s_x + \sin \varphi \tau_0\otimes \sigma_z \otimes s_y).
\end{equation}
We choose $g=\sqrt{2}/10$ and $\varphi=\pi/4$ and calculate the energy spectrum for our 4-layer model on a $24\times24$ lattice, as shown in Fig.~\ref{Fig4} (a). As expected, there exists a pair of zero-energy Majorana modes within both bulk and edge BdG gaps. In Fig.~\ref{Fig4} (b), we further plot the spatial profile of these zero-mode wavefunctions and find them to be localized around the top-right and bottom-left corners. The presence of such Majorana corner modes signals the magnetic-field-induced {\it higher-order topology} \cite{HOTnote} in FeSC thin films. Notably, rotating ${\bf B}_\parallel$ enables ``hopping" of the Majorana corner mode from one corner to another \cite{zhu2018tunable}, allowing for new braiding protocols for Majorana-based topological quantum information processing \cite{Pahomi2020braiding,zhang2020holonomic}.  

{\it Experimental Detection} - To detect the charge-neutral helical Majorana modes, researchers have proposed a variety of measurement schemes including measuring spin or thermal transport signals~\cite{tanaka2009spincurrent}, probing the nonlocal conductance on a superconducting ring~\cite{beri2012nonlocal}, creating a a quantum spin Hall insulator-helical TSC junction~\cite{huang2018helical}, etc. Most of these proposals, however, appear to be technically challenging. Inspired by the field-induced higher-order topology in helical TSCs, we propose a new ``smoking gun" for helical Majorana fermions with a local density of states (LDOS) probe such as scanning tunneling microscopy. Our measurement protocol consists of two steps: 
\begin{enumerate}
	\item[(i)] at zero ${\bf B}_\parallel$, the bulk LDOS shows a hard superconducting gap, while the LDOS spectrum near the sample edge features a ``bell-shaped" peak around the zero bias; 
	\item[(ii)] turning on ${\bf B}_\parallel$ suppresses the original zero-bias peak in the edge LDOS, while a sharp zero-bias peak of LDOS remains localized around the sample corners.
\end{enumerate}
We note that the above step (i) by itself is insufficient to conclude the existence of helical edge Majorana modes, since it may be attributed to edge-localized trivial non-Majorana bound states. However, additionally observing the field-induced Majorana corner modes as described in step (ii) will rule out such a trivial interpretation, providing more definitive evidence for the helical topological nature of the zero-field state.

{\it Discussion} - We have established FeSC thin films as a promising high-temperature platform for gate-controlled helical topological superconductivity. Our recipe for helical TSC consists of three ingredients: the topological Dirac surface states, the intrinsic $s_\pm$ pairing, and the gate-controlled SIA, all of which are experimentally accessible in a thin film of topological FeSC such as FeTe$_{1-x}$Se$_x$, (Li$_{0.84}$Fe$_{0.16}$)OHFeSe, CaKFe$_4$As$_4$, LiFeAs, etc. Using the state-of-the-arts epitaxial growth technique in the FeSC family of materials, our protocol should lead to the observation of helical Majorana modes.   


The convenient electrical tunability of our helical TSC platform enables flexible designs of helical Majorana circuits. For example, it is straightforward to achieve an electric-field-controlled quantum-point-contact geometry in our proposed setup. Such a quantum point contact has been predicted to host exotic properties that are unique to helical Majorana fermions~\cite{asano2010tunneling}. The FeSC thin film also serves as a feasible platform for creating a Josephson junction between a helical TSC and a normal superconductor, which should allow the detection of time-reversal anomaly~\cite{chung2013TRanomaly}.   

{\it Acknowledgment}- R.-X.Z. thanks Xianxin Wu, Lun-Hui Hu, Fan Zhang, Xiao-Qi Sun, Jiabin Yu, and Yang-Zhi Chou for helpful discussions. This work is supported by the Laboratory of Physical Sciences. R.-X.Z. acknowledges the support of a JQI Postdoctoral Fellowship.

\bibliography{HelicalTSC}

\begin{thebibliography}{47}
\bibitem [{HOT()}]{HOTnote}%
  \BibitemOpen
  \href@noop {} {}\bibinfo {note} {Note that the field-induced higher-order
  topology in our system is ``extrinsic", which is protected by the edge Zeeman
  gap instead of the bulk BdG gap}\BibitemShut {NoStop}%
\end{thebibliography}

\onecolumngrid

\subsection{\large Supplemental Material for ``Intrinsic Time-reversal-invariant Topological Superconductivity in Thin Films of Iron-based Superconductors"}

\section{Thickness dependence of topological phase diagram}

In this part, we will study the robustness of helical TSC phase in the topological phase diagram as a function of the thickness of a thin film. 

\begin{figure*}[h]
	\centering
	\includegraphics[width=0.9\textwidth]{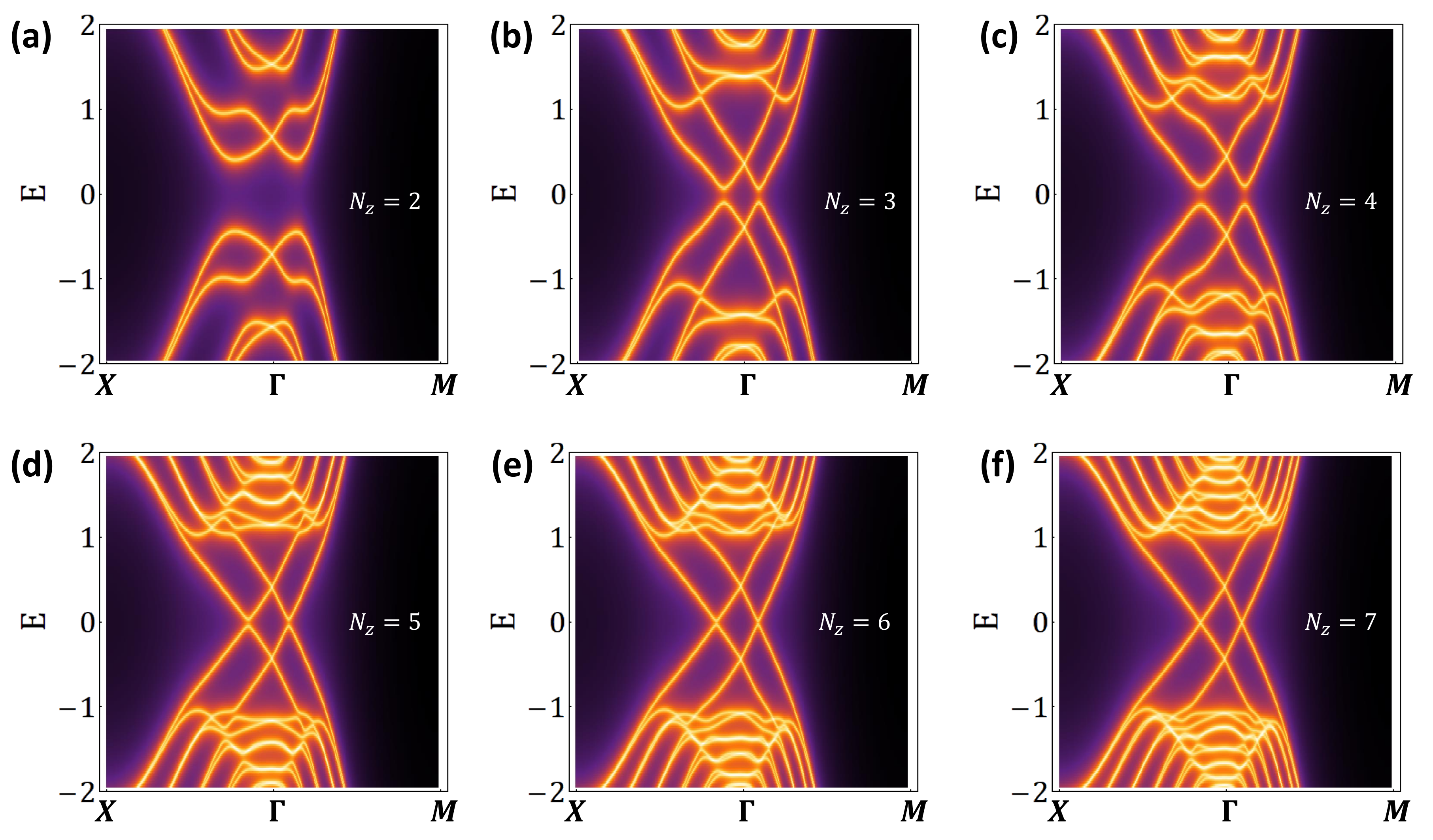}
	\caption{Normal-state energy spectrum of our FeSC thin film model for different $N_z$.}
	\label{fig:band plots}
\end{figure*}

In Fig.~\ref{fig:band plots}, we plot the normal-state energy spectrum of a FeSC thin film as described by our lattice model for $N_z=2,3,...,7$. In particular, for $N_z\geq 6$, we find that hybridization $t$ between the top and bottom surfaces becomes vanishingly small, indicating that the thin film has entered the 3D limit.

\begin{figure*}[h]
	\centering
	\includegraphics[width=0.7\textwidth]{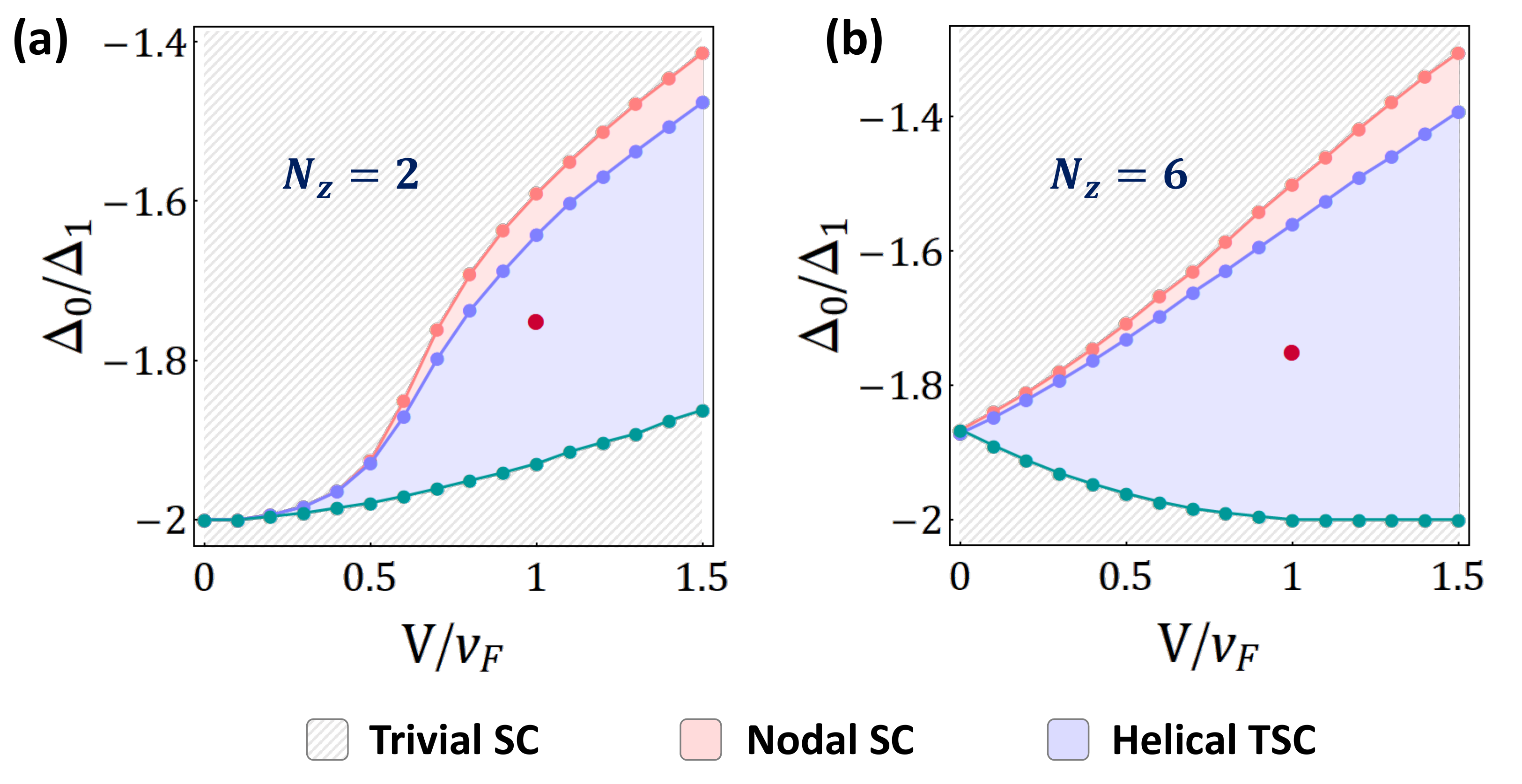}
	\caption{Topological phase diagrams for $N_z=2$ and $N_z=6$ are shown in (a) and (b). The red dot in both (a) and (b) serve as a reference point, which coincides with the one in Fig.~3 in the main text.}
	\label{fig:phase diagram}
\end{figure*}

For our purpose, we further map out the topological phase diagrams for $N_z=2$ and $N_z=6$ in Fig.~\ref{fig:phase diagram} (a) and (b), respectively, to compare with that of $N_z=4$ shown in Fig.~3 (a) of the main text. Based on Fig.~\ref{fig:band plots} (c) and (e), we find the intersurface hybridization $t$ is around 0.2 for $N_z=4$ and almost zero for $N_z=6$. Interestingly, there is no visible difference between the corresponding phase diagrams in Fig.~\ref{fig:phase diagram} (b) [$N_z=6$] and the Fig.~3 (a) in the main text [$N_z=4$]. For $N_z=2$, the system is approaching the 2D limit and experiences a much greater quantum confinement effect when compared to that of the $N_z=4$ case, which manifests in the energy spectrum in Fig.~\ref{fig:band plots} (a). However, we find that the corresponding topological phase diagram in Fig.~\ref{fig:phase diagram} (a) is only quantitatively modified, in comparison to Fig.~3 (a) for $N_z=4$ in the main text. Therefore, we conclude that for $N_z>1$, varying $N_z$ will only quantitatively modify our topological phase diagram, without introducing any qualitative change.

\end{document}